\documentclass[proceedings, preprint]{rmaa}

\usepackage{paralist}

\usepackage{psfrag,color}
\usepackage{times}
\usepackage{graphicx}
\usepackage{ulem}
\usepackage{epsfig}
\usepackage{natbib}

\def\apj{ApJ}
\def\apjl{ApJ}
\def\apjs{ApJS}
\def\ao{Appl.~Opt.}
\def\aap{A\&A}
\def\jcap{J. Cosmology Astropart. Phys.}
\def\mnras{MNRAS}
\def\prd{Phys.~Rev.~D}
\def\nat{Nature}

\newcommand{\msun}{\mbox{$M_\odot$}}

\def\be{\begin{eqnarray}}
\def\ee{\end{eqnarray}}
\def\bi{\begin{itemize}}
\def\ei{\end{itemize}}
\def\lsim{\mathrel{\rlap{\lower3pt\hbox{\hskip1pt$\sim$}}
     \raise1pt\hbox{$<$}}} 
\def\gsim{\mathrel{\rlap{\lower3pt\hbox{\hskip1pt$\sim$}}
     \raise1pt\hbox{$>$}}} 

\newcommand{\bary}{\begin{eqnarray}}
\newcommand{\eary}{\end{eqnarray}}

\SetYear{2013}
\SetConfTitle{Magnetic Fields in the Universe IV}

\title{Neutrino Oscillation from Magnetized Strange Stars} 

\author{
  Nissim Fraija\altaffilmark{1} 
  and Enrique Moreno M\'endez\altaffilmark{2}}

\altaffiltext{1}{Luc Binette-Fundaci\'on UNAM Fellow. Instituto de Astronom\'\i{}a,  UNAM, CU, M\'exico (nifraija@astro.unam.mx).}
\altaffiltext{2}{Instituto de Astronom\'\i{}a,  UNAM, CU, M\'exico (enriquemm@astro.unam.mx).}

\shortauthor{N. Fraija, \& E. Moreno M\'endez}
\shorttitle{RevMexAA(SC) Demo Document}

\listofauthors{N. Fraija \& E. Moreno M\'endez}
\indexauthor{N. Fraija}
\indexauthor{E. Moreno M\'endez}

\abstract{Strange-quark matter (SQM) is a likely candidate of the ground state of nuclear matter. 
Along with many other equations of state (EoSs), SQM seemed to be severely constrained by the recent discoveries of the 1.97 $\msun$ PSR J1614-2230 and the 2.01 $\msun$ PSR J0348+0432. 
However with new, $O(\alpha_c^2)$, perturbative calculations, SQM seems to be able to accommodate masses as large as $\sim 2.75 \msun$.
The literature of SQM stars or strange stars includes estimates of internal magnetic fields as large as $10^{20}$ G, which are unlikely to be formed as they would require  $\sim 10^{57}$ erg to be produced. 
Nonetheless, if strange stars may hold magnetar-strength fields ($10^{15}$ G), their internal fields are likely to reach magnetic fields as large as $10^{17}$ G.
We consider neutrinos with energies of some MeV and oscillation parameters from solar, atmospheric and accelerator experiments. 
We study the possibility of resonant oscillation of neutrinos in strange stars.
}

\resumen{
La materia extra\~na de quarks (SQM) es un candidato probable para el estado base de la materia nuclear.
Junto con otras ecuaciones de estado, la SQM parece estar restringida por los recientes descubrimientos de los pulsares PSR J1614-2230 de 1.97 $\msun$ y PSR J0348+0432 de 2.01$\msun$. 
Nuevos c\'alculos perturbativos, a $O(\alpha_c^2)$, permiten explicar masas de hasta $\sim 2.75 \msun$ para estrellas de SQM.
La literatura de estrellas de SQM o estrellas extra\~nas incluye estimaciones de campos magn\'eticos internos  de hasta $10^{20}$ G, los cuales es poco probable que se formen dado que se requiere del orden de $\sim 10^{57}$ ergs para producirlos.
No obstante, si las estrellas extra\~nas pudiesen tener campos externos de magnetar ($10^{15}$ G), sus campos internos podr\'ian llegar a los $10^{17}$ G.
Consideramos neutrinos con energ\'ias de algunos MeV y par\'ametros de oscilaci\'on obtenidos de mediciones de neutrinos solares, atmosf\'ericos, y aceleradores.
Bajo estas condiciones estudiamos la posibilidad de que exista resonancia en oscilaci\'on de neutrinos en estrellas extra\~nas.
}

\addkeyword{neutrinos}
\addkeyword{stars: evolution}
\addkeyword{stars: magnetic field}
\addkeyword{stars: massive}
\addkeyword{stars: neutron}

\begin{document}
\maketitle

\section{Introduction}

Stars that produce Fe cores end up their lives as gravitational-core-collapse (GCC) supernovae (SNe), most likely driven by neutrinos \citep{1979NuPhA.324..487B,1990RvMP...62..801B}, if the compact object ends below a mass of some $\sim3 \msun$.  
The threshold mass is determined by the nuclear equation of state (EoS) but it could be partially influenced by other factors like centrifugal support on millisecond (ms) pulsars (PSRs) and/or magnetic pressure in, e.g., magnetars \citep[see, e.g.,][for a review]{2010arXiv1012.3208L}.

Recently, the nuclear EoS has been further constrained by the discovery of two massive pulsars, the 1.97-$\msun$ PSR J1614-2230 \citep{2010Natur.467.1081D} and the 2.01-$\msun$ PSR J0348+0432 \citep{2013Sci...340..448A}.
Nonetheless, the nuclear EoS is still not known and many possibilities exist.
Among them, is the possibility that as the pressure and density grow in the inside of a neutron star (NS), the quarks that originally conform a baryon mix with the quarks from neighbouring baryons until all quarks become free and produce quark matter \citep[QM; see, e.g.,][]{1970PThPh..44..291I}.  \citet{Witten:1984rs}, suggested that the Fermi energy of QM could be lowered by allowing an extra degree of freedom in the quarks, namely, another flavor.
Thus, roughly $\sim1/3$ (minus the chemical potential of the new quark flavor), of the quarks are converted to the next lower mass quark, which is the strange quark.  Thus this matter is named strange quark matter (SQM).
According to O($\alpha_C^2$) estimates by \citep{2010PhRvD..81j5021K}, with $\alpha_C$ the strong coupling constant, the maximum mass of SQM stars can be $M_{SQM} \gtrsim 2.75 \msun$.
Following \citet{1984PhRvD..30.2379F}, \citet{Haensel:1986qb},  \citet{1986ApJ...310..261A} and  \citet{1997csnp.conf.....G}, use the MIT-bag model to study the structure and characteristics of strange stars.

\begin{figure}[!t]
  \includegraphics[width=0.8\columnwidth]{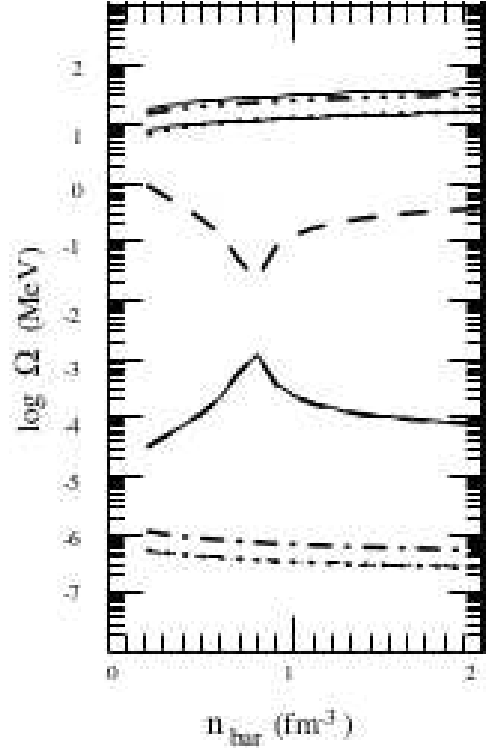}
  \caption{Plot of the baryon density vs. the log of the frequency (in MeV).  The upper five curves are plasma frequencies, whereas the lower five curves are the gyrofrequencies.  The upper continuous line is the total plasma frequency. The Dash-dot-dot-dot line is for the up quark. The slimmer continuous line is for the down quark, and the dotted line is the strange quark.  The dashed curve is for the electrons (lower density) and the positrons (higher density).  The next top-to-down continuous curve is for the total gyrofrequency.  The dashed one is for the electrons and positrons.  The dash-dotted line is for the up quarks. The last two curves are for down and strange quarks.  This plot is for a $10^{12}$ G magnetic field, the gyrofrequency curves get displaced upwards linearly for higher magnetic fields.}
  \label{fig:alfa6}
\end{figure}

Using the MIT-bag model the values for the bag constant, the strong-coupling constant and the mass of the strange quark can be varied.  For values above ~ 0.5 for the strong coupling constant strange quark matter can become negatively charged as the density increases (\citealt{2013arXiv1306.1828M} and Moreno M\'endez \& Page 2013, in preparation).  
Figure~\ref{fig:alfa6} shows how electrons are replaced by positrons once the baryonic density reaches values ~ 0.8 fm-3.   
In such case, a region where e--e+ pairs annihilate into neutrinos may be formed.  
In this scenario neutrinos can be produced in this region even after the strange matter has cooled below 1 MeV. 

SQM stars are bound by QCD and, thus, in principle, they could sustain a rather strong, internal, magnetic field. 
An internal field, perhaps as large as an equipartition one (during GCC), i.e., $\sim 10^{17}$ G could in principle exist.  
Here we will test fields as large as $3 \times 10^{16}$ G.

Now, the properties of neutrinos get modified when they propagate in a strongly magnetized medium.  
Depending on the flavor, a neutrino would feel a different effective potential because electron neutrinos ($\nu_e$) interact with electrons via both neutral and charged currents (CC), whereas muon ($\nu_\mu$) and tau ($\nu_\tau)$ neutrinos interact only via the neutral current (NC). 
This induces a coherent effect in which maximal conversion of $\nu_e$ into $\nu_\mu$ ($\nu_\tau$) takes place even for a small intrinsic mixing angle.  
The resonant conversion of neutrino from one flavor to another due to the medium effect, is well known as the Mikheyev-Smirnov-Wolfenstein effect \citep{1978PhRvD..17.2369W}.
In this work, we study  the propagation and resonant oscillation of thermal neutrinos in strange stars.  
We take into account  the two-neutrino mixing (solar, atmospheric and accelerator parameters) . 
Finally, we discuss our results in the strange stars framework.

\section{Neutrino Production for Large Coupling}
Using the MIT-bag model the values for the bag constant, the strong coupling constant and the mass of the strange quark can be varied. 
For values above ~ 0.5 for the strong coupling constant strange quark matter can become negatively charged as the
density increases.
The figure on the right shows how electrons are replaced by positrons once the baryonic density reaches values ~ 0.8 fm-3. 
In such case, a region where e--e+ pairs annihilate into neutrinos forms. 
In this scenario neutrinos can be produced in this region even after the strange matter has cooled below 1 MeV.

\section{Neutrino  Effective Potential}\label{sec-Justification}

We use the finite temperature field theory formalism to study the effect of heat bath  on the propagation of elementary particles. The effect of magnetic field is taken into account through Schwinger's propertime method \citep{1951PhRv...82..664S,2009PhRvD..80c3009S,2009JCAP...11..024S}.  
The effective potential of a particle is calculated from the real part of its self-energy diagram. The neutrino field equation of motion in a magnetized medium is (Fraija 2013, submitted)
\be
[ {\rlap /k} -\Sigma(k) ] \Psi_L=0,
\label{disneu}
\ee
where the neutrino self-energy operator $\Sigma(k)$ is a Lorentz scalar which depends on the characterized parameters of  the medium, as for instance,  chemical potential, particle density, temperature, magnetic field, etc.  Solving this equation and using the Dirac algebra, the dispersion relation $V_{eff}=k_0-|{\bf k}|$ as a function of Lorentz scalars can be written as 
\be
V_{eff}=b-c\,\cos\phi-a_{\perp}|{\bf k}|\sin^2\phi,
\label{poteff1}
\ee
where $\phi$ is the angle between the neutrino momentum and the magnetic field vector.   Now the Lorentz scalars $a$, $b$ and $c$ which are functions of neutrino
energy, momentum and magnetic field can be calculated from the neutrino self-energy due to charge current and neutral current interaction of neutrino with the background particles.

\subsection{One-loop  neutrino self-energy}

The total one-loop  neutrino self-energy  in a magnetized medium is given by \citep{1998PhRvD..58h5016E, 2014arXiv1401.1581F} 
\be
Re \Sigma_W(k)=R\,[a_{W_\perp} \rlap /k_\perp + b_W \rlap /u + c_W \rlap /b]\,L
\ee
where the Lorentz scalars are given by the following equations
{
\bary
a_{W\perp}&=&-\frac{\sqrt2G_F}{M_W^2}\biggl[
\biggl\{E_{\nu_e}(N_e-\bar{N}_e)+ k_3(N_e^0-\bar{N}_e^0)\biggr\}\nonumber\\
&&+\frac{eB}{2\pi^2}\int^\infty_0 dp_3\sum_{n=0}^\infty(2-\delta_{n,0})\cr
&&\biggl (\frac{m_e^2}{E_n}- \frac{H}{E_n}\biggr) (f_{e,n}+\bar{f}_{e,n})\biggr],
\label{conaw}
\eary
}

{
\bary
b_W&=& \sqrt2 G_F \biggl[\biggl(1+\frac32\frac{m_e^2}{M_W^2}
+\frac{E_{\nu_e}^2}{M_W^2}\biggr)(N_e-\bar{N}_e)\cr
&&+\biggl(\frac{eB}{M_W^2}
+\frac{ E_{\nu_e}k_3}{M_W^2}\biggr)(N_e^0-\bar{N}_e^0)\cr
&&-\frac{eB}{2\pi^2M_W^2} \int^\infty_0 dp_3
\sum_{n=0}^\infty(2-\delta_{n,0})\cr
&&\biggl\{2k_3E_n\delta_{n,0}
+2E_{\nu_e}\biggl(E_n-\frac{m_e^2}{2E_n}\biggr)\biggr\}\cr
&&(f_{e,n}+\bar{f}_{e,n})\biggr]\nonumber
\label{conbw}
\eary
}
and
{
\bary
c_W&=&\sqrt2
G_F\biggl[\biggl(1+\frac12\frac{m_e^2}{M_W^2}-\frac{k_3^2}{M_W^2}\biggr)(N_e^0-\bar{N}_e^0)\cr
&&+\biggl(\frac{eB}{M_W^2}-\frac{E_{\nu_e}k_3}{M_W^2}\biggr)(N_e-\bar{N}_e)\nonumber\\
&&-\frac{eB}{2\pi^2M_W^2} \int^\infty_0
dp_3\sum_{n=0}^\infty(2-\delta_{n,0})\cr
&&\biggl\{2E_{\nu_e}
\biggl(E_n-\frac{m_e^2}{2E_n}\biggr)\delta_{n,0}+2k_3\cr
&&\biggl(E_n-\frac32\frac{m_e^2}{E_n}-
\frac{H}{E_n}\biggr)\biggr\}(f_{e,n}+\bar{f}_{e,n})\biggr].
\label{concw}
\eary
}

In the strong magnetic field approximation, the energy of charged particles is modified confining the particles to the Lowest Landau level ($n=0$). The number density of electrons will become,

\be
N_e^0=\frac{eB}{2\pi^2}\int^\infty_0 dp_3 f_{e,0}
\ee
where 
\be
f(E_{e,0})=\frac{1}{e^{\beta(E_{e,0}-\mu)} +1}
\ee
and the  electron energy in the lowest Landau level is,
\be
E^2_{e,0}=(p^2_3+m_e^2)\,.
\ee
Assuming that  the chemical potentials ($\mu$) of the electrons and positrons are much smaller than their energies ($\mu\leq$E$_e$), the fermion distribution function can be written as a sum  given by
\bary
f(E_{e,0})&=&\frac{1}{e^{\beta(E_{e,0}-\mu)} +1} \nonumber\\
&\approx&\sum^{\infty}_{l=0}(-1)^l e^{-\beta(E_{e,0}-\mu)(l+1)}\, .
\eary

The effective potential is

{
\bary
V_{eff}&=&\frac{\sqrt2 G_F\,m_e^3}{\pi^2}\frac{B}{B_c}\biggl[\sum^\infty_{l=0} (-1)^l \sinh\alpha\,K_1(\sigma)\cr
&& \biggl\{ 1+C_{V_e}+ \frac{m_e^2}{m^2_W}\biggl(\frac32+2\frac{E^2_\nu}{m^2_e} +\frac{B}{B_c}  \biggr)-\cr
&&\biggl(1-C_{Ae} + \frac{m_e^2}{m^2_W}\biggl(\frac12-2\frac{E^2_\nu}{m^2_e} +\frac{B}{B_c}  \biggr) \biggr)\cr
&&\cos\phi  \biggr\}-4\frac{m_e^2}{m^2_W}\frac{E_\nu}{m_e}\sum^\infty_{l=0} (-1)^l \cosh\alpha\cr
&&\biggl\{ \frac34K_0(\sigma)+\frac{K_1(\sigma)}{\sigma}-\frac{K_1(\sigma)}{\sigma}\cos\phi\biggr\}\biggr],\;\;\;\;\;\;
\label{poteff2}
\eary
}
where the critical magnetic field is  $B_c=m^2/e\simeq4.1\times10^{13}$ G,   $K_i$ is the modified Bessel function of integral order i, $\alpha=\beta\mu(l+1)$ and $ \sigma=\beta m_e(l+1)$.

\begin{figure}
\vspace{0.3cm}
{ \centering
\resizebox*{0.45\textwidth}{0.27\textheight}
{\includegraphics{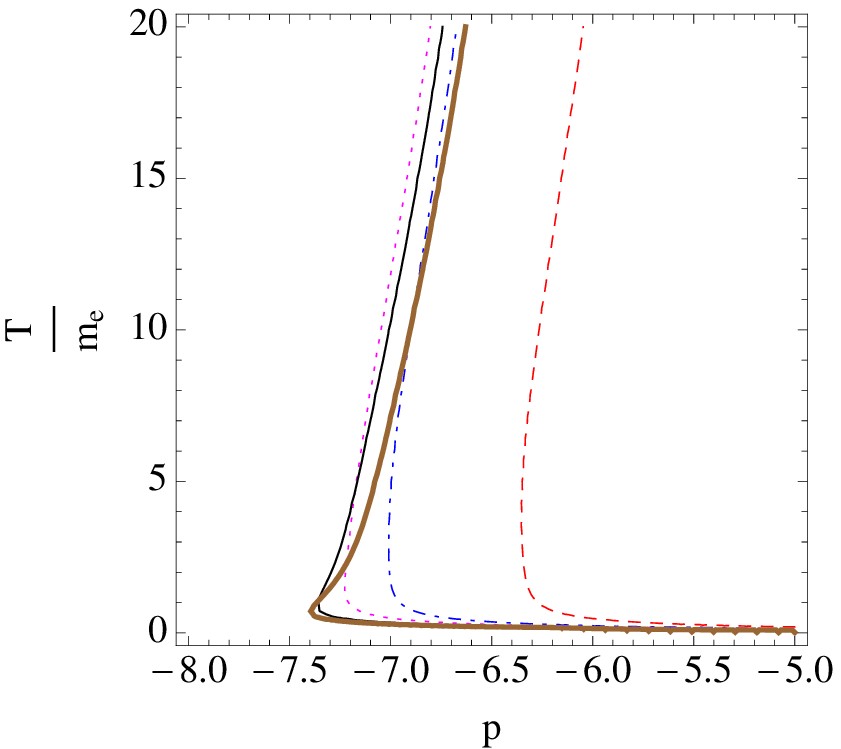}}
\resizebox*{0.45\textwidth}{0.27\textheight}
{\includegraphics{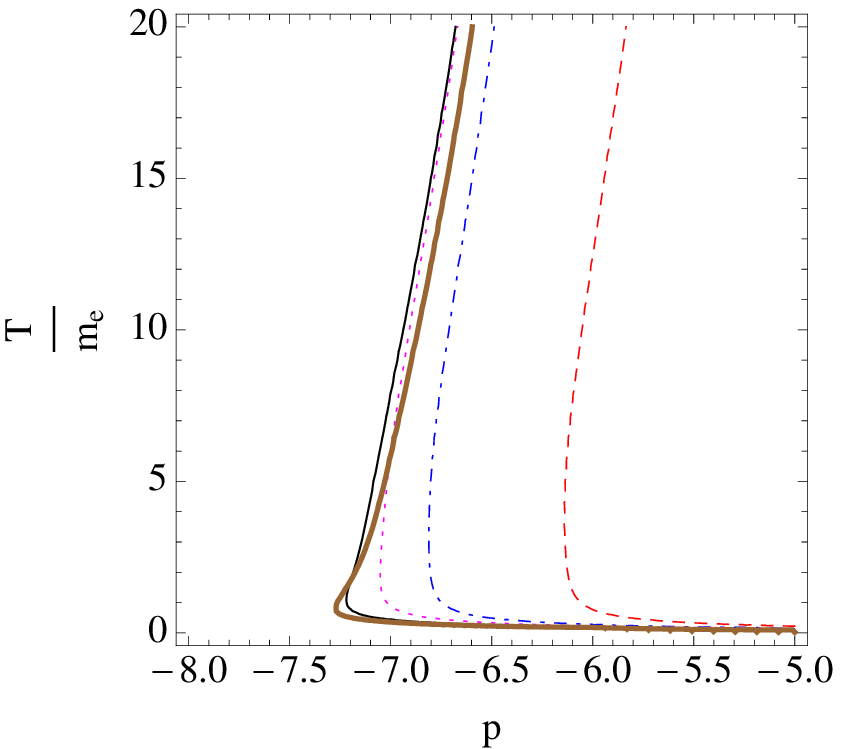}}
\resizebox*{0.45\textwidth}{0.27\textheight}
{\includegraphics{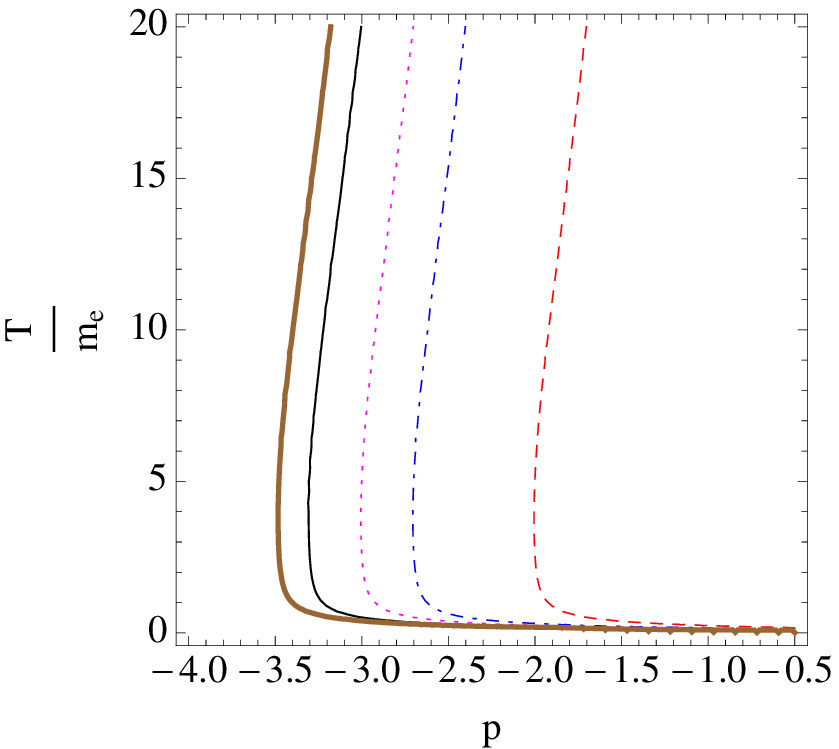}}
}
\caption{Plot of temperature  ($T/m_e$) as a function of chemical potential  ($\mu=m_e10^p$) for which the resonance condition is satisfied.  We have used the best parameters of the two-flavor  solar (top),  atmospheric (middle) and  accelerator (bottom) neutrino oscillation, B=10$^3$ B$_c$ and taken five  different neutrino energies:  E$_\nu=$1 MeV (red dashed line),  E$_\nu=$5 MeV (blue dot-dashed line), E$_\nu=$10 MeV (magenta dotted line), E$_\nu=$20 MeV (black thin-solid line) and E$_\nu=$30 MeV (brown thick-solid line). }
\label{ptime}
\end{figure}

\section{Two-Neutrino Mixing}

Here we consider the neutrino oscillation process $\nu_e\leftrightarrow \nu_{\mu, \tau}$. The evolution equation for the propagation of neutrinos in the above medium is given by \citep{2014MNRAS.437.2187F}
\be
i
{\pmatrix {\dot{\nu}_{e} \cr \dot{\nu}_{\mu}\cr}}
={\pmatrix
{V_{eff}-\Delta \cos 2\theta & \frac{\Delta}{2}\sin 2\theta \cr
\frac{\Delta}{2}\sin 2\theta  & 0\cr}}
{\pmatrix
{\nu_{e} \cr \nu_{\mu}\cr}},
\ee
where $\Delta=\delta m^2/2E_{\nu}$, $V_{eff}$ is the potential difference between $V_{\nu_e}$ and $V_{\nu_{\mu, \tau}}$,    $E_{\nu}$ is the neutrino energy and $\theta$ is the neutrino mixing angle. The conversion probability for the above process at a given time $t$ is given by 
\be
P_{\nu_e\rightarrow {\nu_{\mu}{(\nu_\tau)}}}(t) = 
\frac{\Delta^2 \sin^2 2\theta}{\omega^2}\sin^2\left (\frac{\omega t}{2}\right
),
\label{prob}
\ee
with
\be
\omega=\sqrt{(V_{eff}-\Delta \cos 2\theta)^2+\Delta^2 \sin^2
    2\theta}.
\ee
The effective potential for the above oscillation process is given by eq. (\ref{poteff2}). The oscillation length for the neutrino is given by
\be
L_{osc}=\frac{L_v}{\sqrt{\cos^2 2\theta (1-\frac{V_{eff}}{\Delta \cos 2\theta}
    )^2+\sin^2 2\theta}},
\label{osclength}
\ee
where $L_v=2\pi/\Delta$ is the vacuum oscillation length.   The resonance length can be written as
\be
L_{res}=\frac{L_v}{\sin 2\theta} 
\ee
to obtain the previous equation,  we applied the resonance condition given by
\be
V_{eff} -  \frac{\delta m^2}{2E_{\nu}} \cos 2\theta = 0.
\label{reso}
\ee
\acknowledgments
NF gratefully acknowledges a Luc Binette-Fundaci\'on UNAM Posdoctoral Fellowship. EMM was supported by a CONACyT fellowship and projects CB-2007/83254 and CB-2008/101958.  This research has made use of NASAs Astrophysics Data System as well as arXiv. 
\section{conclusions}
We have plotted the resonance condition for  neutrino oscillation in strongly magnetized strange stars. 
We have taken into account  the  temperature is in the range of 1- 10 MeV and magnetic field B=10$^3 B_c$. 
We have used  effective potential $V_{eff}$ up to order M$_W^4$, the best parameters for the two- solar \citep{2011arXiv1109.0763S}, atmospheric \citep{2011PhRvL.107x1801A} and accelerator \citep{1996PhRvL..77.3082A, 1998PhRvL..81.1774A} neutrinos and the neutrino energies ($E_\nu = 1$, 5, 10, 20 and 30 MeV). 
From the plots in fig.~\ref{ptime} we can see that the chemical potential for electrons from accelerator parameters is largest and the one from solar parameters the smallest.
Another important charcteristic from fig.~\ref{ptime} is that the temperature is degenerated for solar and atmospheric parameters unlike the case for accelerator parameters.
Further strange star observables (for instance leptonic asymmetry, minimum baryon load, etc.) will be estimated in Fraija \& Moreno M\'endez (2013, in preparation).


\end{document}